\documentclass[twocolumn,a4paper,superscriptaddress]{revtex4-1}
\usepackage[utf8]{inputenc}
\usepackage[english]{babel}
\usepackage{amsmath}
\usepackage{amssymb}
\usepackage{amsfonts}
\usepackage{color}
\usepackage{graphicx}
\usepackage{subfigure}
\usepackage{epsfig}
\usepackage{pxfonts}
\begin{document}
\title{Elastic enhancement factor as a quantum chaos probe}
\author{Yaroslav A. Kharkov}
\affiliation{Novosibirsk State University - Novosibirsk, Russia}
\affiliation{Budker Institute of Nuclear Physics of SB RAS -
Novosibirsk, Russia}
\author{Valentin V. Sokolov}
\email{V.V.Sokolov@inp.nsk.su}
\affiliation{Budker Institute of Nuclear Physics of SB RAS -
Novosibirsk, Russia}
\affiliation{Novosibirsk Technical University - Novosibirsk,
Russia}

\pacs{05.45.-v, 24.30.+b}
\begin{abstract}
Recent development of the resonance scattering theory with a
transient from the regular to chaotic internal dynamics inspires 
renewed interest to the problem of the elastic enhancement
phenomenon. We reexamine the question what the experimentally
observed value of the elastic enhancement factor can tell us
about the character of dynamics of the intermediate system.
Noting first a remarkable connection of this factor with the time 
delays variance in the case of the standard Gaussian ensembles we 
then prove the universal nature of such a relation. This reduces 
our problem to that of calculation of the Dyson's binary form
factor in the whole transition region. By the example of systems 
with no time-reversal symmetry we then demonstrate that the
enhancement can serve as a measure of the degree of internal
chaos.
\end{abstract}

\maketitle

Excess of probabilities of elastic processes over inelastic ones 
is a common feature of the resonance compound nuclear reactions, 
electron transport through quantum dots, where it manifests
itself as the weak localization effect, or, at last, transmission 
of electromagnetic waves through microwave cavities. This
phenomenon, that is characterized quantitatively by the {\it
elastic enhancement factor} i.e. the typical ratio $F$ of elastic
and inelastic cross sections, repeatedly attracted attention for 
decades \cite{Collection, Lewenkopf, Pluhar, Verbaarshot,
Fyodorov}. Based on the random matrix theory \cite{Mehta}
universal formalism that allows of uniform treatment of all
resonance phenomena of such a kind has been worked out in the
seminal paper \cite{Zirnbauer}. The scattering $M\times M$ matrix 
that describes the resonance processes is expressed as
\begin{equation}\label{S}
S(E) = I - i A^{\dag}\frac{1}{E-H+\frac{i}{2}A\,A^{\dag}}\,A\,.
\end{equation}
The Hamiltonian matrix $H$ that describes dynamics of the
originally closed intermediate system is supposed to belong to an 
ensemble of random matrices. This system gets excited and then
decays after a while because its $N\gg 1$ eigenstates are
connected to $M$ open channels via random \cite{Lehmann}
rectangular $N\times M$ matrix $A$. The formula (\ref{S}) can be 
naturally interpreted \cite{Rotter, Sokolov} (see also
\cite{Sommers, Mitchell} and references therein) within the
concept of open systems whose internal dynamics is described by a 
non-Hermitian effective Hamiltonian $\mathcal{H} = H -
\frac{i}{2}W$ where the connection to channels results in the
anti-Hermitian contribution proportional to the matrix
$W=A\,A^{\dag}$. The complex eigenvalues of the effective
Hamiltonian specify positions and widths of the resonances.\\

All quantities of the physical meaning are obtained by averaging 
$\langle ...\rangle$ over two independent ensembles of matrices
$H$ and $A$ \cite{Lehmann}. Specifically, the averaged scattering 
matrix $\langle S^{a b}(E)\rangle=\langle S^{a
a}(E)\rangle\,\delta^{a b}$ fixes the transmission coefficients
$T^a=1-|\langle S^{a a}\rangle|^2$ that measure the part of the
flow that spends essential time in the internal region. The scale 
of the energy dependence of the mean scattering matrix as well as 
the transmission coefficients is large and their energy
variations can be neglected within the energy interval where the 
$N$ resonance states of interest are situated.\\

In the most interesting case of large number $M\gg 1$ of
scattering channels (for the sake of simplicity we suppose that
all these channels are statistically equivalent and have,
therefore, identical transmission coefficients $T$) the
enhancement factor $F$ consists of two  contributions of quite
different nature. The first of them, $1+\delta_{1 \beta}$,
depends only on the symmetry class $\beta=1$ (preserved
time-reversal (T) invariance) or $\beta=2$ (broken T-invariance) 
of the corresponding ensembles of the random effective
Hamiltonians ${\cal H}$ \cite{Lewenkopf, Lehmann, Pluhar}. No
enhancement originates from this contribution in the case of
Unitary ensemble of complex Hermitian matrices. The second
contribution is regulated by the ratio $\eta=t_H/t_W=M T$
("openness") of two characteristic times. One of them, the
Heisenberg time
\begin{equation}\label{t_H}
t_H=\frac{2\pi}{d}=-2\Im\left\langle\mathrm
{Tr}\,G(E+i0)\right\rangle\,,
\end{equation}
where $G(E+i0)=\left(E+i0-H\right)^{-1}$ is the resolvent of the 
Hamiltonian $H$, characterizes the internal motion and is defined 
by its mean level spacing $d$.
Similarly \cite{Dittes, Savin}, the dwell time
$t_W=\frac{1}{T}\langle Q\rangle_{A\rightarrow\, 0}$, where
$\langle Q\rangle=-\frac{2}{M}\Im\langle\mathrm
{Tr}\frac{1}{E-\mathcal {H}}\rangle$ is the mean delay time,
establishes the time scale of the open system in terms of the
mean Wigner delay time \cite{Wigner, Smith}. The dwell time is
the time the incoming particle spends in the internal region. The 
inverse quantity $\Gamma_W=1/t_W$ is nothing else than the well-known 
Weisskopf width \cite{Weisskopf}. Only if the openness is
small enough, $\eta\ll 1$, so that the dwell time appreciably
exceeds the Heisenberg time, $t_H\ll t_W\,$, the incoming
particle has enough time to recognize the discreteness of the
internal spectrum and therefore to perceive spectral
fluctuations. Then an additional contribution appears
\cite{Verbaarshot, Fyodorov} and the enhancement factor takes
finally the form
\begin{equation}\label{F_lim}
F^{(\beta)}(\eta)=1+\delta_{\beta 1}+\eta\,\int_0^{\infty} d s
e^{-\eta s} \,\left[1-B_2^{(\beta)}(s)\right]\,\\
\end{equation}
where $B_2^{(\beta)}(s)$ is the Dyson's spectral binary form
factor \cite{Mehta} belonging to the symmetry class $\beta$.
Comparing this expression with the delay time two-point
correlation function
\begin{equation}\label{K}
\begin{array}{c}
K^{(\beta)}(\epsilon)=\frac{\langle Q(\frac{\epsilon}{2})
Q(-\frac{\epsilon}{2})\rangle}{\langle Q^2\rangle}-1\\
=\int_0^{\infty} d s\,e^{-\eta s}\left[1-B^{(\beta)}_2(s)\right]
\cos\left(2\pi s\rho\epsilon\right)\\
\end{array}
\end{equation}
calculated for the case $\beta=1$ in \cite{Savin} (later on,
similar calculation has been performed also in the case $\beta=2$ 
\cite{Fyodorov+}) we arrive at the following remarkable relation 
between the enhancement factor and variance of the time delays:
\begin{equation}\label{F_var Q}
\begin{array}{c}
F^{(\beta)}(\eta)=1+\delta_{\beta 1}+\eta\,var Q(\eta)\\
=2+\delta_{\beta 1}-\eta\,\int_0^{\infty} d s\, e^{-\eta s}
\,B_2^{(\beta)}(s)\,.
\end{array}
\end{equation}\\

A new aspect of the old problem of the elastic enhancement has
been recently evoked in ref.\cite{Celardo} (see also earlier
semiclassical consideration in \cite{Baranger}) where
manifestations of transition from regular to chaotic internal
motion has been investigated in the framework of the resonance
scattering theory. The character of this motion is controlled by 
a particle interaction parameter $\kappa$. The dynamics of the
intermediate system can therefore be described by some transient 
matrix ensemble. It has been, in particular, numerically
discovered that the elastic enhancement factor is quite sensitive 
to the strength of the interaction. This fact suggests that the
enhancement factor can serve as an indicator of the degree of the 
internal chaoticity. In what follows we investigate this relation 
analytically.\\

Generally, the short range universal fluctuations of scattering
amplitudes are described by the (connected) $S$-matrix two-point 
auto-correlation function \cite{Zirnbauer}
\begin{equation}\label{S-S}
 C^{abcd}(\epsilon) = \langle S^{a b}(E+\epsilon/2) S^{c
 d\,*}(E-\epsilon/2)\rangle_{conn}\,.
\end{equation}
While carrying out the ensemble averaging we, suppose throughout 
this Letter the decay amplitudes $A_n^c$ to be uncorrelated
Gaussian random quantities
\begin{equation}
\langle A_n^a A_m^{b\,*}\rangle_A =
\frac{\gamma}{N}\delta_{nm}\delta^{ab}\,.
\end{equation}
This assumption is supported by the reasons of so-called
"geometrical chaos" that have been argued in ref.
\cite{Zelevinsky}. In the case of T-invariant systems $\beta=1$
these amplitudes are real, $A_m^{b\,*}=A_m^{b}$.\\

For the elastic enhancement factor the formula (\ref{S-S}) gives  
$F = C^{aaaa}(0)/C^{abab}(0)\,$. We start ensemble averaging with 
that over the amplitudes $A$ keeping the internal Hamiltonian $H$ 
diagonal. It is convenient (though not necessary) to use
supersymmetric integral representation. Then $A$-averaging can be 
fulfilled exactly whereupon the saddle point method can be used. 
In such a way we receive first of all
\begin{equation}\label{mean_S}
\langle S^{ab}(E)\rangle = \delta^{a
b}\left\langle\frac{1-i\frac{\gamma}{2} g(E)}{1+i\frac{\gamma}{2}
g(E)}\right\rangle_P\,,
\end{equation}
where the function
$g(E)\equiv\frac{1}{N}\langle\mathrm{Tr\frac{1}{E-{\cal H}}}
\rangle_A$ satisfies the equation (m=M/N)
\begin{equation}\label{g(E)}
 g(E) = \frac{1}{N}\sum_{n=1}^{N}\left[E-E_n
 +\frac{\frac{i}{2}m\gamma} {1+\frac{i}{2}\gamma
 g(E)}\right]^{-1}\,.
\end{equation}

The subscript $P$ in the r.h.s. of eq. (\ref{mean_S}) implies
averaging over all energy levels of the internal system with the 
joint probability distribution $P(\{E\}|\kappa)$; $\{E\}\equiv
E_1, E_2,...E_N$ at a given value of the chaoticity parameter
$\kappa$. Depending on this parameter, the distribution
$P(\{E\}|\kappa)$ changes from Poissonian distribution of fully
independent levels ($\kappa=0$) to that of highly correlated
levels what is typical of the Gaussian ensembles
($\kappa=\infty$). We assume also that the mean level density
does not depend on $\kappa$ at all.\\

In the limit of weak coupling to continuum $\gamma\rightarrow 0$ 
we are interested in, the approximate solution of eq.
(\ref{g(E)}) reads
\begin{equation}\label{g_1(E)}
g(E)\approx
\frac{1}{N}\mathrm{Tr}\,G\left(E+\frac{i}{2}m\gamma\right)\,.
\end{equation}
We suppose below that the ratio $m$ is also small. Then the mean 
scattering matrix reduces \cite{Sokolov, Sokolov+} to
\begin{equation}\label{app_mean_S}
\langle S^{ab}(E)\rangle = \delta^{a b}\frac{1-i\frac{\gamma}{2}
\left\langle g(E)\right\rangle_P}{1+i\frac{\gamma}{2}\left\langle
g(E)\right\rangle_P}=\frac{1-x}{1+x}\,.
\end{equation}
In accordance with the conventional practice, we neglected the
long range energy dependence of the mean $S$-matrix elements and 
set $E=0$. (We will do the same in all later calculations.) The
measuring the degree of resonance overlapping parameter
$x=\frac{\pi\gamma}{2N\,d}$ (where $d$ is the mean level spacing) 
should be small in the case of our interest so that the
corresponding transmission coefficients equal $T\approx 4\,x$
and, correspondingly, the openness is $\eta=4M\,x=2\pi
m\frac{\gamma}{d}\,$.\\

The tensor structure of the correlation function $(\ref{S-S})$
\begin{equation}\label{S-S-T}
 C^{abcd}(\epsilon) =
 F_1^{(\beta)}(\epsilon)\delta^{ab}\delta^{cd} +
 F_2^{(\beta)}(\epsilon)(\delta^{ac}\delta^{bd} +
 \delta_{1\beta}\delta^{ad}\delta^{bc})\,
\end{equation}
follows from the T-invariance properties and rotational
invariance in the channel space. The superscript $\beta$ marks
now the symmetry class of the limiting ($\kappa=\infty$) matrix
ensemble. The enhancement factor reads therefore
\begin{equation}\label{F}
F^{(\beta)} = 1+\delta_{\beta 1} +
\left[F_1^{(\beta)}(\epsilon)\Big/F_2^{(\beta)}(\epsilon)\right]_{\epsilon=0}\,.
\end{equation}
Now, in the leading order with respect to $\gamma$ the
$A$-averaging results in
\begin{equation}\label{eq:F^12}
\begin{array}{l}
 F_1^{(\beta)}(N,M,\gamma|\kappa) =
 \frac{\gamma^2}{N^2}\left\langle \mathrm{Tr}\,
 G(\frac{i}{2}m\gamma) \mathrm{Tr}\,G^\dag(\frac{i}{2}m\gamma)
 \right\rangle_{P,\,conn}\nonumber\\
 F_2^{(\beta)}(N,M,\gamma|\kappa) =
 \frac{\gamma^2}{N^2}\left\langle \mathrm{Tr}\,
 G(\frac{i}{2}m\gamma)
 G^\dag(\frac{i}{2}m\gamma)\right\rangle_{P}\,.
 \end{array}
\end{equation}
Subsequent $P$-averaging is straightforward and leads to the
expression
\begin{equation}\label{F^1/F^2}
F_1^{(\beta)}\Big/F_2^{(\beta)}=\eta \int_0^{\infty} d
s\,e^{-\eta s}\left[1-B_2^{(\beta)}(s|\kappa)\right]\,.
\end{equation}
Note that this ratio depends after all only on two parameters: on 
the openness $\eta$ of the internal system and on the degree of
chaoticity $\kappa$ of its dynamics. Finally, we arrive at the
expression
\begin{equation}\label{F_var Q_k}
F^{(\beta)}(\eta|\kappa)=2+\delta_{\beta 1}-\eta\int_0^{\infty} d
s\,e^{-\eta s} B_2^{(\beta)}(s|\kappa)
\end{equation}
that extends the relations (\ref{F_lim}, \ref{F_var Q})) to the
case of arbitrary value of the chaoticity parameter.\\

The found result reduces the problem posed above to that of
calculating the binary form factor $B_2^{(\beta)}(s|\kappa)$ in
the whole transient region $0\leqslant\kappa<\infty$. The issue
of transition from regular to chaotic dynamics has been attacked 
not once by different authors (see \cite{Guhr98} and references
therein). The total solution has been found by now only in the
case of the systems with broken time-reversal symmetry
\cite{Pandey, Shapiro}. The method used in \cite{Shapiro} is the 
most convenient for our purpose. These authors has used the
Brezin-Hikami's approach \cite{Brezin96} that allows of direct
calculating the binary form factor we need. Below we restrict
ourselves to the case $\beta=2$ and will skip this
superscript.\\

The following two properties of the considered binary form factor 
are obvious from the very beginning: $B_2(s|0)=0$ and
$B_2(s|\infty)=(1-s)\theta(1-s)$, \cite{Mehta}.
In the intermediate region the form factor is given by
\cite{Shapiro}
\begin{equation}\label{eq:B_2(s|k)}
\begin{array}{l}
 B_2(s|\kappa)=B_2(s|\infty)\\
 -\frac{2}{\pi}\int_{-1}^1 dy\frac{[2y\sqrt{s}+1]\sqrt{1-y^2}}{s
 + 2y\sqrt{s} + 1} e^{-\kappa s(s + 2y\sqrt{s} + 1)}\,.
\end{array}
\end{equation}
In fact, only the even part of the integrand contributes. The
enhancement factor is entirely expressed via the function
\begin{equation}\label{Psi}
\begin{array}{c}
\Psi(\eta|\kappa)=\eta\int_0^{\infty}d
s\,e^{-\kappa\,s(s+1)-s\eta}\,\,
\frac{I_1(2\kappa\, s^{3/2})}{\kappa\, s^{3/2}}\\
=\eta\int_0^{\infty}d
s\,e^{-\kappa\,s(s+1)+2\kappa\,s^{3/2}-s\eta}\,
\Xi(2\kappa\,s^{3/2})\\
\end{array}
\end{equation}
where $I_1(x)$ stands for the modified Bessel function and the
function
$\Xi(2\kappa\,s^{3/2})=e^{-2\kappa\,s^{3/2}}\frac{I_1(2\kappa\,s^{3/2})}{\kappa\,s^{3/2}}$ 
decreases monotonously from one to zero when the argument
$2\kappa\,s^{3/2}$ grows. It is easy to check that
$\Psi(\eta|0)=1$, $\,\Psi(\eta|\infty)=0$ and
$\Psi(\eta|\kappa)>0$ in between.\\

There are two equivalent representations of the enhancement
factor:
\begin{equation}\label{F_k small}
F(\eta|\kappa)=1+\Psi(\eta|\kappa)+\eta\frac{\partial^2}
{\partial\eta^2}\left[\frac{1}{\eta}\int_0^{\kappa}d
\kappa'\Psi(\eta|\kappa')\right]\,,
\end{equation}
and
\begin{equation}\label{F_k large}
\begin{array}{c}
F(\eta|\kappa)=1+(1-e^{-\eta})\left/\eta\right.\\
\,\,\,\,\,\,\\
+\Psi(\eta|\kappa)-\eta\frac{\partial^2}{\partial\eta^2}
\left[\frac{1}{\eta}\int_{\kappa}^{\infty}d
\kappa'\Psi(\eta|\kappa')\right]\,.
\end{array}
\end{equation}
In particular, the first formula gives immediately
$F(\eta|0)\equiv 2$ when the second one reduces to the GUE result 
$F_{GUE}=1+(1-e^{-\eta})/\eta\approx 2-\frac{1}{2}\eta+...\,$.
More than that, it can be shown with the aid of
eq. (\ref{F_k large}) that the slope at the point $\eta=0$ is
universal:
\begin{equation}\label{Slope}
\frac{\partial F(\eta|\kappa\neq 0)}{\partial
\eta}\Big|_{\eta=0}\equiv -\frac{1}{2}\,.
\end{equation}
Indeed, at any nonzero $\kappa$ contributions of the two last
terms cancel each other when $\eta\rightarrow 0$ with accuracy
better than $\eta$.\\

Although one cannot obtain any exact explicit analytical formula, 
a number of approximate expressions can be derived from eqs.
(\ref{F_k small}, \ref{F_k large}). At that eq.(\ref{F_k small}) 
is useful when the internal chaoticity is weak whereas the second 
form is more convenient if the internal dynamics is close to
chaotic. Behavior of the function $\Psi$ depends on interrelation 
of the two competing parameters $\kappa$ and $\eta$. If the first 
of them is small enough and the second one is kept finite we can 
take into account only few leading terms in the Bessel function
power series. Corresponding contributions are expressed already
in the terms of known transcendental functions. After that there 
are two possibilities: either expand these functions into power
series with respect to the parameter $\kappa$ or make expansion
over inverse powers of the parameter $\eta$. In the first case
coefficients of the $\kappa$-expansion are polynomials in
$1/\eta$; in the second case those of the $1/\eta$-expansion are 
polynomials in $\kappa$. The two found in such a way expansions
do not perfectly coincide. Nevertheless they match with certain
accuracy that can be improved by taking into account a larger
number of contributions. Substituting finally the estimated in
such a way function $\Psi$ into eq. (\ref{F_k small}) we arrive
at
\begin{equation}\label{k_small, eta_fix}
F(\eta|\kappa) = 2 - \frac{\kappa}{\eta} +
\frac{(6+\eta)\kappa^2}{\eta^3}  - \frac{(60 + \eta (20 +
\eta))\kappa^3}{\eta^5}+...\,.
\end{equation}

On the other hand, if the parameter of chaoticity is large,
$\kappa\gg 1$, calculations become appreciably simpler and
integration in eq. (\ref{Psi}) can be carried out with the help
of the Laplace method. At that, it is convenient to utilize the
presentation given in the second line of eq. (\ref{Psi}).
Generally speaking, the exponential factor in the integrand has
two maxima in the points $s_0=0$ and $s_1=1/8\left(5-4\eta/\kappa 
+3\sqrt{1-8\eta/\kappa}\right)\,.$ In the first of them the whole
integrand equals one. Then $\Psi(\eta|\kappa)\approx
\frac{\eta}{\eta+\kappa}$ and the contribution of the vicinity of
the point $s_0$ in the enhancement factor is easily found to be
\begin{equation}\label{k_large}
F(\eta|\kappa)\approx
1+\frac{1-e^{-\eta}}{\eta}+\frac{\eta}{\eta+\kappa}-\frac{\eta}{(\eta+\kappa)^2}\,.
\end{equation}
As to the second point $s_1$, the maximum of the exponential
reaches its largest  possible value one when $\eta=0$, goes
rapidly down with growing $\eta$ and, after passing the
inflection point $s_i=9/16$, disappears finally when $\eta$
exceeds $\kappa/8\,.$ But even in the most interesting case
$\kappa\gg\eta\gtrsim 1$ contribution of the vicinity of the
second maximum remains negligible. Indeed, opposite to the hight 
of the maximum neither its position $s_1\approx 1$, nor its width 
$\Delta s\approx 0.4$ noticeably depend on $\eta\,.$ Therefore
the slow varying factor can be estimated as $\Xi(2\kappa
s^{3/2})\approx\Xi(2\kappa s_1^{3/2})\approx
1\left/\sqrt{2\pi}\,\kappa^{3/2}\right.\ll 1\,.$\\

The Fig.1 illustrates variations of the elastic enhancement
factor depending on the increasing openness $\eta$ and parameter 
chaoticity $\kappa$. In particular, the domains of validity of
our approximations are demonstrated. At any given chaoticity
$\kappa$ the factor $F(\eta|\kappa)$ decreases with the universal 
initial slope $-1/2$ starting from the maximal value 2 up to some 
value $\eta_c(\kappa)$ where this factor reaches minimum
$F_{min}$ so that
\begin{equation}\label{eta_c}
\frac{\partial F(\eta|\kappa)}{\partial \eta}
\Big|_{\eta=\eta_c(\kappa)}=0.
\end{equation}

The level correlations at the chosen value of the chaoticity
parameter $\kappa$ are too weak to be resolved when openness
$\eta$ exceeds the value $\eta_c(\kappa)$ and the enhancement
factor returns to value 2 typical of the system with regular
dynamics. In this way the enhancement factor conveys information 
on the degree of internal chaos. Finally, the required chaoticity 
parameter $\kappa(F_{min})$ is found as the root of the equation 
$F\left(\eta_c(\kappa)|\kappa\right)=F_{min}\,.$\\

\begin{figure}\label{fig:Enh_Factor}
{\includegraphics[width=0.5\textwidth]{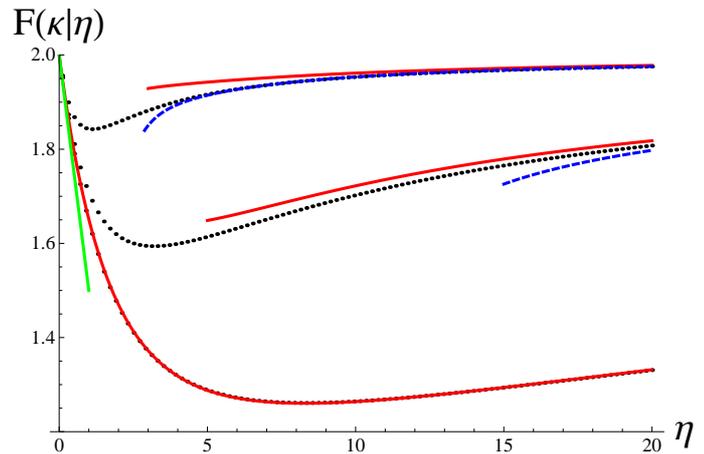}}
\caption{(color online)
Enhancement factor $F(\eta|\kappa)$ as a function of $\eta$ for
three fixed values (from top to bottom) $\kappa = (0.5, 5, 50)$.
Black points - exact numerical results, blue lines (dashed) -
small $\kappa$ approximation (\ref{k_small, eta_fix}), red lines
(full) - large $\kappa$ approximation (\ref{k_large}); the
straight green (gray) line shows the universal slope at the point
$\eta=0$.}
\end{figure}

Though we do not know the explicit form of the binary form factor 
$B^{(\beta=1)}_2(s|\kappa)$ that describes considered transition 
in the case of $T$-invariant systems, there is no doubt that,
qualitatively, the behavior of the enhancement factor should be
similar. It is worth mentioning that the evolution of this factor 
under transition from perfectly chaotic systems with broken
$T$-invariance to $T$-invariant ones can easily be followed with 
the aid of the method developed in \cite{Fyodorov++}.\\

The further quite nontrivial development reported in
\cite{Brezin03} opens opportunity of deriving such a form factor 
for T-invariant systems also. Corresponding results will be
published elsewhere.

\vspace{0.6cm}
{\bf Summary} We have considered the dependence of the elastic
enhancement factor on the degree of chaoticity of the internal
part of an open resonance system. A general relation of this
factor with the variance of time delays has been established for 
arbitrary degree of chaoticity. By this, the task is reduced to
the search for the transient binary form factor. Generally, the
enhancement factor $F(\eta|\kappa)$ depends on both the
chaoticity parameter $\kappa$ and the openness $\eta$ the latter 
being the ratio $\eta=t_H/t_W$ of two characteristic times: the
dwell time $t_W$ and the Heisenberg time $t_H$. Here the time
$t_H=\frac{2\pi}{d}$ is the time that is needed to resolve the
pattern of spectral fluctuations in the system with Hamiltonian
$H$ when the time $t_W=1/\Gamma_W$ is that the incoming particle 
spends in average inside this system. Only if this particle is
trapped inside for sufficiently long time it can carry
information on the internal chaos. Otherwise the difference
between regular and chaotic internal dynamics cannot be
resolved.\\

In this Letter, the problem posed has been thoroughly studied
numerically and analytically in the case of systems with no
time-reversal symmetry. We showed in particular that the slope
$\frac{\partial F(\eta| \kappa\neq 0)}{\partial
\eta}\Big|_{\eta=0}=-\frac{1}{2}$ remains invariable for
arbitrary degree of internal chaos. The recovery of the maximal
value of $F(\eta|\kappa)$ when the openness $\eta$ exceeds some
value $\eta_c(\kappa)$ that is clearly seen in the Fig.1 is in
perfect agreement with the physical argumentation stated in the
previous paragraph.

\vspace{0.3cm}
We are very grateful to V.G. Zelevinsky and V.F. Dmitriev for
useful discussions. This work is supported by the Ministry of
Education and Science of the Russian Federation. Y. Kharkov
acknowledges financial support by the ``Dynasty'' foundation,
by the Government of Russian Federation (grant 11.G34.31.0035),
by the foundation "Leading Scientific Schools of Russia"
(grant 6885.2010.2) and the Russian Foundation for basic Research (grant 12-01-00943-a). V. Sokolov appreciates financial support from the federal program "personnel of innovational Russia" (grant 14.740.11.0082) as well as countenance by the RAS Joint scientific program "Nonlinear dynamics and Solitons" .

\end{document}